\RequirePackage{snapshot}
\documentclass[lettersize,journal]{IEEEtran}

\usepackage{amsmath,amsfonts, amssymb, stmaryrd}
\usepackage{algorithmic}
\usepackage{algorithm}
\usepackage{array}
\usepackage[caption=false,font=normalsize,labelfont=sf,textfont=sf]{subfig}
\usepackage{textcomp}
\usepackage{stfloats}
\usepackage{url}
\usepackage{verbatim}
\usepackage{graphicx}
\usepackage{cite}
\usepackage{xcolor}
\usepackage{tikz}
\usepackage{hyperref}
\usepackage{ulem}

\usetikzlibrary{positioning, fit, shapes.geometric, arrows, decorations.pathreplacing}

\tikzstyle{box} = [rectangle, text centered]
\tikzstyle{arrow} = [thick,->,>=stealth]
\tikzstyle{textbox} = [rectangle, align=center, draw=black, fill={rgb, 255:red, 249; green, 231; blue, 202}, draw={rgb, 255:red, 242; green, 202; blue, 140}, thick, minimum height=1cm,minimum width=2cm]
\tikzstyle{textellipse} = [ellipse, align=center, draw=black, thick, minimum height=1.5cm, minimum width=3cm, fill=white]

\begin{document}
	
	\title{Finding Incompatible Blocks \\for Reliable JPEG Steganalysis}
	\author{Etienne Levecque, Jan Butora, Patrick Bas\thanks{The authors are with the University of Lille, CNRS, Centrale Lille, UMR 9189 CRIStAL Lille, France. Email: firstname.lastname@univ-lille.fr}}
	
	\markboth{}{}
	
	\maketitle
	
	\begin{abstract}
		This article presents a refined notion of incompatible JPEG images for a quality factor of 100. It can detect the presence of steganographic schemes embedding in DCT coefficients.
		We show that, within the JPEG pipeline, the combination of the DCT transform with the quantization function can map several blocks in the pixel domain to the same block in the DCT domain. However, not every DCT block can be obtained: we call those blocks \textit{incompatible}. In particular, incompatibility can happen when DCT coefficients are manually modified to embed a message.
		We show that the problem of distinguishing compatible blocks from incompatible ones is an inverse problem with or without solution and we propose two different methods to solve it. The first one is heuristic-based, fast to find a solution if it exists. The second is formulated as an Integer Linear Programming problem and can detect incompatible blocks only for a specific DCT transform in a reasonable amount of time.
		We show that the probability for a block to become incompatible only relies on the number of modifications.
		Finally, using the heuristic algorithm we can derive a Likelihood Ratio Test depending on the number of compatible blocks per image to perform steganalysis. We simulate the result of this test and show that it outperforms a deep learning detector e-SRNet for every payload between 0.001 and 0.01 bpp by using only 10\% of the blocks from $\bf 256\times 256$ images. A Selection-Channel-Aware version of the test is even more powerful and outperforms e-SRNet while using only 1\% of the blocks.
	\end{abstract}
	
	\begin{IEEEkeywords}
		Steganography, Steganalysis, JPEG, Compatibility, Compatibility attacks, Reliability
	\end{IEEEkeywords}
	
	\section{Introduction}
	\label{sec:1_intro}
	In steganography, the main objective is to hide a message inside an innocuous media such as an image or a video called a Cover media. To be secure, the steganography message needs to be undetectable from someone performing steganalysis, i.e. trying to distinguish Cover from Stego media by analyzing them. This cat-and-mouse game between the steganographer and the steganalyst is very dependent on a lot of parameters, such as the media format used to hide the message, the payload size, or the embedding scheme used.
	
	For this paper, we focus on the JPEG format, which is still one of the most common compression schemes used for digital images because of its relatively good compression properties and its very widespread across different software and hardware implementations. We target the class of embedding schemes that modify the quantized DCT coefficients of the JPEG image. This represents a wide variety of embedding schemes such as JSTEG~\cite{westfeld1999attacks,kodovsky2010quantitative} developed in 1995, or popular academic schemes such as JUniward~\cite{holub2014universal} or UERD~\cite{guo2015using}.
	\begin{figure}[t]
		\centering
		\includegraphics[width=\linewidth]{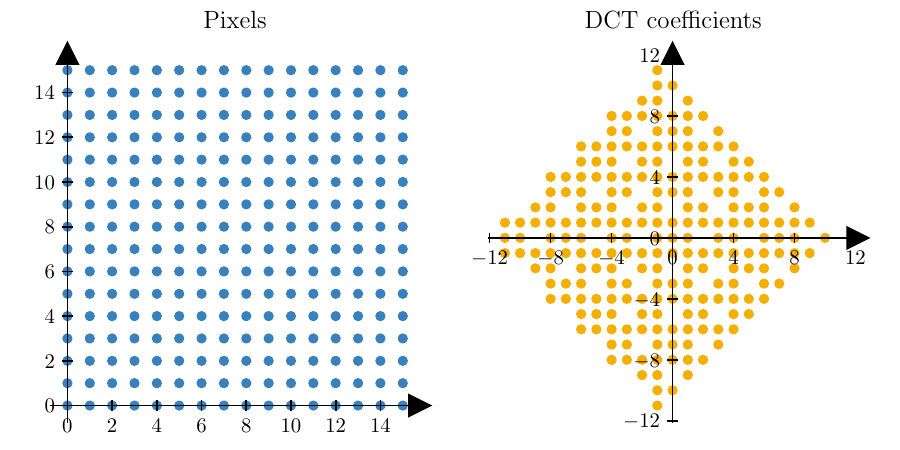}
		\caption{Toy illustration of a 2D JPEG compression illustrating the incompatibility attack. Every block of $1\times 2$ pixels (left figure) is mapped to a point in the DCT space (right figure) using a 2-point DCT algorithm~\eqref{eq:dct}. We can see that the compression is not surjective ,meaning that some blocks in the DCT space (represented by the holes) do not have any antecedent in the pixel domain. On the other hand, some pixel blocks are mapped to the same DCT block during compression. In standard JPEG blocks with 64 dimensions, we observe the same property. The attack exploits the fact that during embedding, some incompatible blocks are created.}
		\label{fig:banner}
	\end{figure}
	
	In this paper, we present a method that exploits a property of JPEG images compressed using a known compression pipeline, which can be used to detect messages for all steganography algorithms altering JPEG quantized DCT coefficients. The main idea of this property is rather simple: given a (compression) pipeline and an image coded in the JPEG format, we want to know if the coded image is an output of this pipeline. This problem, illustrated in Fig.~\ref{fig:banner}, can be seen as an inverse problem because if we can find an input image of the pipeline that gives as output the observed image, then we can conclude that the output image is {\it compatible} with the pipeline. On the contrary, if no input generates the output image, then it is {\it incompatible} with this pipeline. In our case, it means that the image under scrutiny has been tampered with by the embedding algorithm and that it is therefore incompatible with the compression scheme.
	
	The incompatibility we study in this paper is related to the JPEG compression pipeline but not only since it can also be used by any coding scheme which uses the DCT transform and quantized DCT coefficients with quantization steps close to or equal to $1$. This means that this method can be potentially used on coding schemes such as HEIC for still pictures, or the H26x class of coding schemes for moving pictures.
	
	Note that from a steganalysis perspective, detecting an incompatible image is very interesting since in this case the steganalyst, Eve, does not have to face potential false positives. She is sure that the image is Stego if she has knowledge of the compression pipeline and can find at least one incompatible block in the image. Being able to detect incompatible images is consequently related to performing reliable steganalysis.
	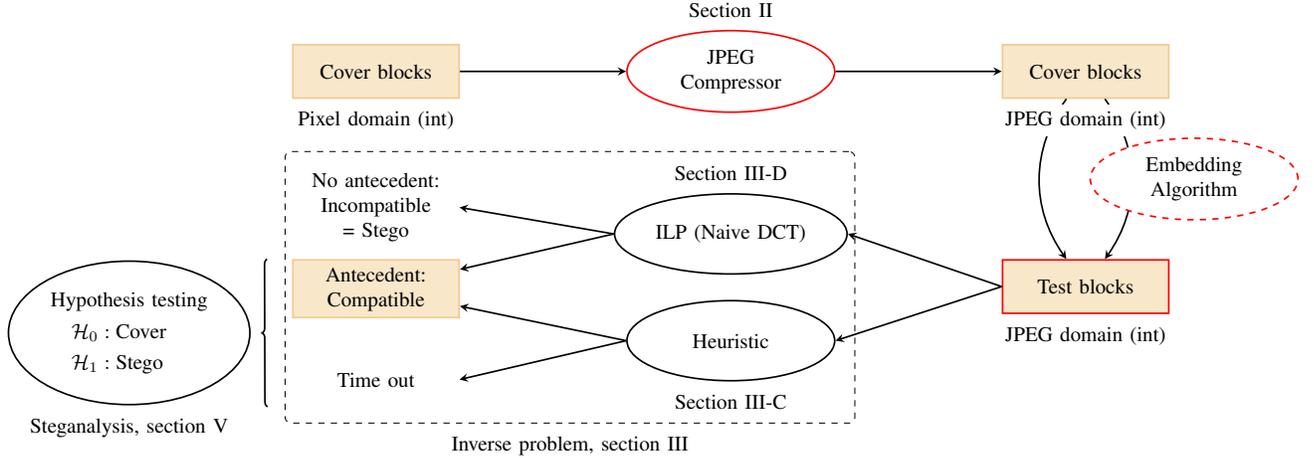
\begin{figure*}[t]
		\centering
		\resizebox{1\textwidth}{!}
		{
			\def\nodedistance{1}
			\def\xsep{1.4}
			\def\ysep{0.9}
			\tikzstyle{box} = [rectangle, text centered]
			\tikzstyle{arrow} = [thick,->,>=stealth]
			\tikzstyle{textbox} = [rectangle, align=center, minimum height=1* \ysep cm,minimum width=2* \xsep cm]
			\tikzstyle{coloredtextbox} = [rectangle, align=center, draw=black, fill={rgb, 255:red, 249; green, 231; blue, 202}, draw={rgb, 255:red, 242; green, 202; blue, 140}, thick, minimum height=1* \ysep cm,minimum width=2* \xsep cm]
			\tikzstyle{textellipse} = [ellipse, align=center, draw=black, thick, minimum height=1.5* \ysep cm, minimum width=2.5* \xsep cm, fill=white]
			\begin{tikzpicture}[node distance=\nodedistance cm]
			\node[coloredtextbox] (pixelcoverblocks) {Cover blocks};
			\node[textellipse, draw=red, right=2*\xsep cm of pixelcoverblocks] (jpegcompressor) {JPEG\\Compressor};
			\node[coloredtextbox, right=2*\xsep cm of jpegcompressor] (dctcoverblocks) {Cover blocks};
			\node[coloredtextbox, draw=red, below=3 * \ysep cm of dctcoverblocks] (testblocks) {Test blocks};
			\node[textellipse, below=3.5*\ysep cm of jpegcompressor] (heuristic) {Heuristic};
			\node[textellipse, below=1.5*\ysep cm of jpegcompressor] (ilp) {ILP (Naive DCT)};
			\node[coloredtextbox, below=3 * \ysep cm of pixelcoverblocks] (compatible) {Antecedent:\\ Compatible};
			\node[textbox, below=1.25 * \ysep cm of pixelcoverblocks] (incompatible) {No antecedent:\\ Incompatible\\= Stego};
			\node[textbox, below=4.75 * \ysep cm of pixelcoverblocks] (timeout) {Time out};
			
			\draw[decoration={brace, mirror, raise=0.3*\xsep cm}, decorate, thick](compatible.north west) -- node[textellipse, left=0.5*\xsep cm] (hyptest) {$
				\begin{aligned}
				&\text{Hypothesis testing}\\
				&\quad \mathcal{H}_0: \text{Cover}\\
				&\quad \mathcal{H}_1: \text{Stego}\\
				\end{aligned}$} (timeout.south west);

			\draw[arrow] (dctcoverblocks) to[bend right=35] (testblocks);
			\draw[arrow] (dctcoverblocks) to[bend left=35] node[textellipse, xshift=0.75*\xsep cm, draw=red,dashed] {Embedding\\Algorithm} (testblocks);

			\node[below=0.1*\ysep cm of pixelcoverblocks, fill=white] (pixeldomain) {Pixel domain (int)};
			\node[below=0.1*\ysep cm of dctcoverblocks, fill=white] (jpegdomaincover) {JPEG domain (int)};
			\node[below=0.1*\ysep cm of testblocks, fill=white] (jpegdomaintest) {JPEG domain (int)};
			
			\node[above=0.1*\ysep cm of jpegcompressor, fill=white] (section2) {Section \ref{sec:jpeg}};
			\node[below=0.1*\ysep cm of heuristic, fill=white] (section3c) {Section \ref{sec:heuristic}};
			\node[above=0.1*\ysep cm of ilp, fill=white] (section3d) {Section \ref{sec:ilp}};
			\node[below=0.1*\ysep cm of hyptest, fill=white] (section4) {Steganalysis, section \ref{sec:steganalysis}};
			
			\node[dashed,rounded corners=.1cm, draw, fit=(heuristic) (ilp) (compatible) (incompatible) (timeout) (section3c) (section3d)] (inverse) {};
			
			\node[below=0.1*\ysep cm of inverse, fill=white] (inverseproblem) {Inverse problem, section \ref{sec:incomp}};
			
			\draw[arrow] (pixelcoverblocks.east) -- (jpegcompressor.west);
			\draw[arrow] (jpegcompressor.east) -- (dctcoverblocks.west);
			
			\draw[arrow] (testblocks.west) -- (heuristic.east);
			\draw[arrow] (testblocks.west) -- (ilp.east);
			\draw[arrow] (ilp.west) -- (incompatible.east);
			\draw[arrow] (ilp.west) -- (compatible);
			\draw[arrow] (heuristic.west) -- (compatible);
			\draw[arrow] (heuristic.west) -- (timeout.east);

			\end{tikzpicture}
		}
		\vspace{-2em}
		\caption{Framework of our method for finding incompatible blocks of a JPEG compressor. Red contours highlight the materials available to Eve to perform the steganalysis study.}
		\label{fig:framework}
	\end{figure*}
	
	This notion of {\it incompatibility} was one of the first strategies in steganalysis and was introduced in 2001 in Fridrich {\it et al.} seminal paper~\cite{fridrich2001steganalysis} when the message is hidden in the decompressed JPEG image. In this case, the observed image is a pixel image formerly compressed in JPEG with some potential modifications on pixel values due to steganographic embedding. The studied pipeline is consequently the JPEG decompression scheme. The authors recompress the image and to detect the embedding, they use on one side the difference signal between the test image in the pixel format and on the other side the recompressed-decompressed image.
	
	Several years later, the Reverse JPEG Compatibility Attack (RJCA) was introduced in Butora and Fridrich's paper~\cite{butora2019reverse} for the case where the message is embedded into the DCT domain of high-quality JPEGs, {\it i.e.} for quality factors equal or close to 100. For this attack, the output image is a JPEG-compressed image and the pipeline is the JPEG compression. The authors use the rounding errors of the decompressed image as a reference signal to detect the embedding, either by simply computing the variance of the signal (which increases after embedding) or using deep neural networks such as \mbox{e-SRNet}~\cite{butora2019reverse}.
	
	Note that incompatibility has also been used for digital forensics to detect tempering operations. The incompatibility between different types of interpolation used during the demosaicking process was proposed by Kirchner and Böhme~\cite{kirchner2009synthesis} in 2009. A more general approach to detect resampled signals and incompatibilities related to interpolation pipelines was proposed by Vázquez-Padín {\it et al.}~\cite{vazquez2013set} using set-membership approaches to estimate the resampling factor when the sampling pipeline - here the interpolation kernel - is known.
	
	A common assumption in steganalysis is called Selection-Channel-Aware (SCA) and was used for the first time in Denemark \textit{et al.} paper~\cite{denemark2014selection} to improve an existing detector. This assumption supposes that the agent performing steganalysis knows the probabilities with which the cover elements have been modified during embedding. In our case, those elements are DCT coefficients and this assumption can also be used to select blocks that are more likely to be modified and thus more likely to be incompatible.
	
	The outline of this contribution is illustrated in Fig.~\ref{fig:framework} where the connections with the different sections of the paper are also highlighted.
	\begin{itemize}
		\item Because the incompatibility stems from the JPEG compression pipeline, different types of compressors are presented in section~\ref{sec:jpeg} with different implementations of the DCT transform together with different rounding functions. The compressor associated with the potential use of an embedding scheme produces blocks that are either Cover or Stego.
		\item The steganalyst analyses the produced image DCT blocks and tries to solve an inverse problem, defined in section~\ref{sec:incomp}, to find antecedents of the test blocks in the pixel domain. Two methods are presented to find an antecedent. The first one uses a heuristic function which is minimized to find compatible blocks. The second one consists in solving an Integer Linear Programming (ILP) problem which can be easily formulated for one specific JPEG scheme and which can be used to detect incompatible blocks. From the heuristic, it is possible to design a timing attack which can be seen as an extension of the one presented in~\cite{levecque:hal-04098582} for a different optimization process and different JPEG compressors.
		\item The timing attack can be finally cast as a hypothesis test based on the number of compatible or timed-out blocks found in an image. Different strategies to select a subset of the blocks, and thus be more efficient, are proposed and the detection performance is evaluated and compared with SOTA deep-learning methods in section~\ref{sec:results}.
	\end{itemize}
	
	Note that to perform steganalysis, the only materials, that Eve needs to have, represented in red in Fig.~\ref{fig:framework}, are:
	
	\begin{itemize}
		\item The knowledge of the JPEG compressor which can be determined by forensic analysis or assumed to be known according to the Kerckhoffs' principle,
		\item Sample blocks of the test image to analyze,
		\item Optionally, the knowledge of the embedding scheme can be used to select the best candidate blocks using an SCA approach.
	\end{itemize}
	
	This paper has some connections with~\cite{levecque:hal-04098582} presenting the main feature of compatibility and its potential connections with timing attacks. However, there are also numerous differences, among them are: the fast heuristic approach presented in section~\ref{sec:heuristic} and used to provide results in section~\ref{sec:statistics} and section~\ref{sec:results}, the design of a Likelihood Ratio Test in section~\ref{sec:likelihood_ratio_test}, the option to have a Selection Channel Aware test using an appropriate prior, and the comparison with the SOTA in deep-learning steganalysis in section~\ref{subsec:results}.
	
	Unlike the original work dealing with the RJCA~\cite{butora2019reverse}, the proposed method is not statistical in nature but exploits the hard constraints of the JPEG format and consequently does not suffer from false positives. Additionally, we show that the proposed Likelihood Ratio Test outperforms in section~\ref{subsec:results} the best deep learning detector used for the RJCA.
	
	All our code is available on our git repository.\footnote{https://gitlab.cristal.univ-lille.fr/elevecqu/incompatible-jpeg-blocks}
	
	\section{JPEG Compression}
	\label{sec:jpeg}
	The goal of this section is to formalize mathematically the JPEG compression pipeline. This formalization will allow us to search for possible antecedents of JPEG blocks. Notations are first defined, then the different steps of the libjpeg implementation~\cite{libjpeg}, which is the most popular implementation of JPEG compression, are presented.
	
	\subsection{Notation and mathematical formulation}
	\textbf{Notations}: Every standard letter represents a scalar and bold letters can represent a vector or a matrix. Most of the values are two-dimensional but can be broadcasted to one dimension by flattening the object. Therefore, for simplicity reasons, multiplication and division of matrix and vector are element-wise operations on the flattened objects. 
	The rounding function to the nearest integer toward infinity is written $\left[\cdot\right]$. For example $[0.5] = [1.4] = 1$ and $[-0.5] = -1$. The quantization operator for some quantization steps stored in a matrix $\mathbf{Q}$ is defined by $[\cdot]_{\mathbf{Q}} = [\frac{\cdot}{\mathbf{Q}}]$. Finally, we define $[\cdot]_{\mathbf{Q}}^{[0;255]}$ as the operation of rounding and then clipping the values to the range $[0;255]$.
	
	\textbf{JPEG compression}: We consider an image divided into non-overlapping blocks of size $8\times8$ pixels. We suppose for simplicity that the image size can be divided by 8. The JPEG compression is applied independently on every block. Let $\mathbf{x}$ be one of those pixel blocks. Its DCT coefficients $\mathbf{c}$ can be defined using the following notation:
	\begin{alignat*}{3}
	\mathbf{d} &= f_{\text{DCT}}(\mathbf{x}) \qquad &&\text{Floating point coefficients,}\\
	\mathbf{c} &= [\mathbf{d}]_{\mathbf{Q}} \qquad &&\text{Integer coefficients,}
	\end{alignat*}
	where $f_{\text{DCT}}$ refers to the forward Discrete Cosine Transform function and $\mathbf{Q}$ is the quantization table of the same size as a block, which depends on the quality factor (QF) of the compression. A high QF means small quantization steps (at QF100, the quantization steps are all equal to $1$).
	
	\textbf{JPEG decompression}: The decompression process is almost symmetric and can be defined with the following notation:
	\begin{alignat*}{3}
	\mathbf{d'} &= \mathbf{Q} \times \mathbf{c} \qquad &&\text{Integer coefficients,}\\
	\mathbf{y}\, &= f_{\text{IDCT}}(\mathbf{d'}) \qquad &&\text{Floating point pixels,}\\
	\mathbf{x'} &= [\mathbf{y}]^{[0;255]} \qquad &&\text{Integer pixels clipped.}
	\end{alignat*}
	
	\begin{figure}[t]
		\def\nodedistance{2.15}
		\begin{tikzpicture}[node distance=\nodedistance cm]
		\node (pos1) [xshift=-1*\nodedistance cm, yshift=-0.5*\nodedistance cm] {};
		\node (pos2) [xshift=3*\nodedistance cm, yshift=-0.5* \nodedistance cm] {};
		\node (pos3) [xshift=0.5 * \nodedistance cm, yshift=0.75 * \nodedistance cm] {};
		\node (pos4) [xshift=0.5* \nodedistance cm, yshift=-1.5 * \nodedistance cm] {};
		\node (x) {$\mathbf{x}$};
		\node (d) [below of=x] {$\mathbf{d}$};
		\node (c) [right of=d] {$\mathbf{c}$};
		\node (dprim) [right of=c] {$\mathbf{d}'$};
		\node (y) [above of=dprim] {$\mathbf{y}$};
		\node (xprim) [right of=y] {$\mathbf{x'}$};
		
		\draw[loosely dotted] (pos1) -- node[below, xshift=-1.75*\nodedistance cm, yshift=-0.12*\nodedistance cm] {\rotatebox{90}{DCT}} node[above, xshift=-1.75*\nodedistance cm, yshift=0.12*\nodedistance cm] {\rotatebox{90}{Pixel}}  (pos2);
		
		\draw[loosely dotted] (pos3) -- node[left, yshift=0.75*\nodedistance cm, xshift=-0.12* \nodedistance cm] {Unknown} node[right, yshift=0.75*\nodedistance cm, xshift=0.12*\nodedistance cm] {Known}(pos4);
		
		\draw [arrow] (x.south) -- node [left, fill=white, xshift=-0.05*\nodedistance cm]  {\rotatebox{-90}{$f_{\text{DCT}}(\cdot)$}} (d.north);
		\draw [arrow] (d.east) -- node [above, fill=white] {$[\cdot]_{\mathbf{Q}}$} (c.west);
		\draw [arrow] (c.east) -- node [above, fill=white] {$\mathbf{Q} \times \cdot$} (dprim.west);
		\draw [arrow] (dprim.north) -- node [right, fill=white, , xshift=0.05*\nodedistance cm]  {\rotatebox{90}{$f_{\text{IDCT}}(\cdot)$}} (y.south);
		\draw [arrow] (y.east) -- node [above, fill=white] {$[\cdot]^{[0;255]}$} (xprim.west);
		\end{tikzpicture}
		\vspace{-2.5em}
		\caption{JPEG pipeline illustrating the notations. The separation unknown/known refers to what the steganalyzer knows or does not.}
		\label{fig:1_jpeg}
	\end{figure}
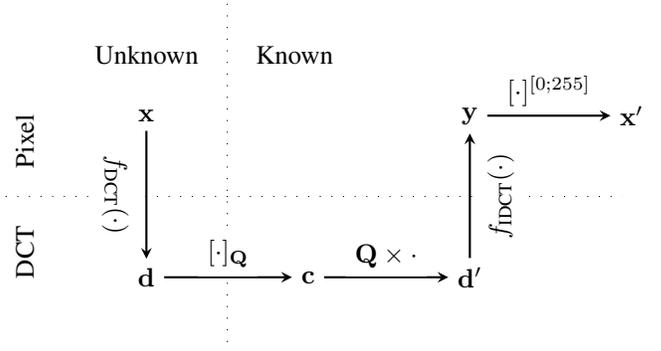
	
	This pipeline is illustrated in figure \ref{fig:1_jpeg} with two separations. The horizontal separation refers to the domain of the objects: either DCT (or frequency) or pixel (or spatial). The vertical separation divides the objects into unknown or known from the steganalyst's point of view.
	
	\textbf{Rounding errors}: The main reason why the JPEG compression is not perfectly reversible is because of the lossy rounding operation in the DCT domain. Indeed, the integer DCT coefficients are a close approximation of the exact floating point value, and therefore a bit of information is lost making the whole pipeline irreversible.
	
	We define the DCT rounding error as well as the spatial rounding error which occurs after decompression to obtain integer pixel values, and the compression error:
	
	\begin{equation}
	\begin{aligned}
	\mathbf{u} &= \mathbf{c} - \mathbf{d} \qquad &&\text{DCT rounding error,}\\
	\mathbf{e} &= \mathbf{x'} - \mathbf{y} \qquad &&\text{Spatial rounding error,}\\
	\mathbf{k} &= \mathbf{x'} - \mathbf{x} \qquad &&\text{Compression error.}
	\end{aligned}
	\label{eq:errors}
	\end{equation}
	
	\subsection{Details on implementations (libjpeg)}
	Libjpeg~\cite{libjpeg} was implemented in 1991 by the Independent JPEG Group as an open-source software written in C. It rapidly became a building brick for high-level languages to use JPEG compression. The JPEG standards are technical specifications so the JPEG implementation can differ depending on the application.
	That is also the case in the libjpeg library, which evolved through different versions~\cite{benes2022knowyl}. This paragraph aims to highlight the most important parameters used for our work.
	
	\textbf{Channels}: This paper will consider 3 different types of image channels. Gray-scale is a single-channel image, the compression is directly performed on every block. RGB-JPEG is a 3-channel image but every channel is compressed like a gray-scale image. YCbCr is a transformation of the color into luminance (Y) and chrominance  (Cb and Cr) components.
	
	Although the transformation is linear, it is followed by a non-linear quantization toward integers. The luminance and chrominance quantization tables also differ in this setting because the human eye is more sensible to the luminance component. Therefore the chrominance components are optionally subsampled and are quantized more than the luminance component, which is subject to the same compression as a gray-scale image.
	
	\textbf{DCT transforms}: The change of domain is the most time-consuming operation during the compression. That is why there exist multiple implementations depending on the quality or speed requirements of the user. There are 3 main implementations in libjpeg: \texttt{float}, \texttt{islow}, and \texttt{ifast}. The \texttt{float} and \texttt{islow} use the same algorithm with floating point and integer values respectively. The \texttt{islow} is therefore faster than \texttt{float} but less precise. The \texttt{ifast} uses a different algorithm on integer values and therefore is even faster than \texttt{islow}. However, the quality of the result can be damaged by the lack of precision especially for high QF values. Both algorithms are presented in~\cite{pennebaker_jpeg_1992}. Since this work targets high QF, \texttt{ifast} will be ignored. Finally, only \texttt{islow} and naive DCT will be considered. The naive DCT is the mathematical definition using a matrix product that is very close to \texttt{float} but can sometimes differ due to operation ordering~\cite{schlogl2023causes}.

	\textbf{Rounding functions}: The quantization step after the DCT makes the use of a rounding function mandatory to transform floating point values to integers. There exist numerous ways to round a floating point to an integer, including truncation (rounding towards zero)~\cite{agarwal2020photo}.
	
	In libjpeg, the rounding function used is the half-up toward infinity implemented with bit shifts. The paper uses the same one. Images compressed with the truncate function are ignored during compression for the \texttt{float} and \texttt{islow} implementations. Note that a rounding function can produce different results depending on the processor of the computer and its instructions~\cite{schlogl2023causes}.
	
	Now that the different mathematical entities have been presented we can give more insights on the definition and the detection of compatible and incompatible blocks.
	
	\section{Incompatibility in JPEG blocks}
	\label{sec:incomp}
	The incompatibility problem is defined in this section. A paragraph is dedicated to a 2-dimensional toy example which gives a geometrical interpretation of the problem. We use two formulations to derive two algorithms to prove that a block is compatible or not. The first one is based on a heuristic algorithm. The second one relies on stronger assumptions to derive an Integer Linear Programming problem that can be solved by numerical solvers.
	
	\subsection{Definition}
	
	A block $\mathbf{c}$ is compatible with a forward DCT function $f_{\text{DCT}}$ and a quantization function $[\cdot]_{\mathbf{Q}}$ if there exists an integer block $\mathbf{x}$ such that:
	\begin{equation}
	\label{eq:inverse_prob}
	\mathbf{c} = \left[f_{\text{DCT}}(\mathbf{x})\right]_{\mathbf{Q}}.
	\end{equation}
	
	In other words, $\mathbf{c}$ is compatible if it has an antecedent given a compression pipeline. On the contrary, if a block of DCT coefficients does not have any antecedent, then it is defined as incompatible.
	
	The notion of compatibility is very simple but the existence of incompatible blocks is not trivial. It comes from the non-surjectivity of the pipeline as depicted in Fig.~\ref{fig:banner}: all pixel blocks $\mathbf{x}$ have an {\it image} by definition but in the space of DCT coefficients, not all blocks $\mathbf{c}$ have an {\it antecedent}. This is mainly due to the quantization function which is not surjective but also due to several implementations of the DCT which are not always perfectly reversible because of precision errors.
	
	In order to hide a message inside the DCT coefficients, the steganographer needs to change the coefficients of a compatible DCT block $\mathbf{c}$ (it was obtained by compressing a pixel block $\mathbf{x}$), but the result $\mathbf{c'}$ after modification is not guaranteed to be compatible and can be exploited by the steganalyst.
	
	\subsection{Why do we have incompatible blocks?}
	
	\begin{figure}[t]
		\centering
		\includegraphics[width=1.\linewidth]{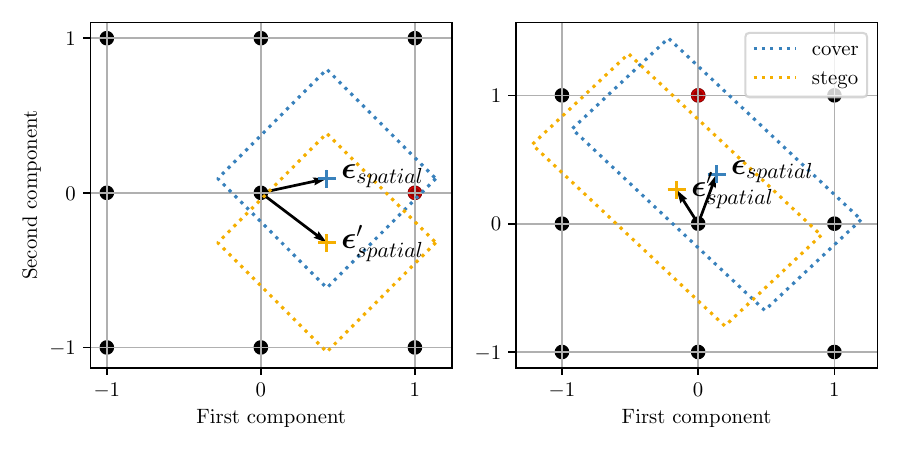}
		\vspace{-2.5em}
		\caption{Toy illustration in 2D. Starting from a random block of $1\times 2$ pixels compressed with the 2-point DCT~\eqref{eq:dct}, we extract $\mathbf{e}$ the rounding error in the spatial domain and $\mathbf{e'}$ the rounding error after embedding a message in the DCT coefficients. The dotted contours represent areas where an integer coordinate can be a solution to problem~\eqref{eq:inverse_prob}. If no integer coordinate falls into this rectangle - for example the orange rectangle on the left plot - it means that there is no solution to the inverse problem and therefore the block is incompatible. On the left, the quantization values are $[1,1]$ (2D equivalent of QF100), and on the right $[1,2]$ (2D equivalent of a QF lower than 100). The red dot is the compression error $\mathbf{k}$.}
		\label{fig:example}
	\end{figure}
	
	In this paragraph, we present a simple toy example with blocks of $1\times 2$ pixels depicted in Fig.~\ref{fig:example}. The same framework using  only $1\times 2$ dimensions was used in Fig.~\ref{fig:banner}. Since we observe the same properties as in the real $8\times8$ dimensional problem, the purpose of this illustration is to give the reader intuitions regarding the creation of incompatible blocks.
	
	Let $\mathbf{x} \in \left\llbracket0,255\right\rrbracket^2$ be a random block of pixels. We use the naive mathematical definition of the DCT as follows:
	\begin{equation}
		\label{eq:dct}
		X_{k,l} = w_k w_l\sum_{i = 0}^{R-1}\sum_{j=0}^{C-1}x_{i,j} \cos\left[\frac{\pi}{R}\left(i + \frac{1}{2}\right)k\right] \cos\left[\frac{\pi}{C}\left(j + \frac{1}{2}\right)l\right].
	\end{equation}
	With $w_k=\frac{1}{\sqrt{R}}$ if $k=0$ and $w_k = \sqrt{ \frac{2}{R}}$ otherwise. $w_l$ is defined similarly with $C$ instead of $R$.
	And generalize it to our blocks of two pixels with $R = 2$ and $C = 1$. We keep the same notations present in Fig.~\ref{fig:1_jpeg} and, only for this section, we consider that $\mathbf{c}$ is a cover DCT block and $\mathbf{c'} = \mathbf{c} + \mathbf{m}$ is the same block with a modification $\mathbf{m}$ that contains some message embedded in it. Note that in the case of a hidden message, the cover block $\mathbf{c}$ is unknown, only the block $\mathbf{c'}$ is observed by Eve.
	
	If we extract the different errors defined in equation~\eqref{eq:errors} we now have two spatial rounding errors $\mathbf{e}$ and $\mathbf{e'}$. Those errors are shown in Fig.~\ref{fig:example} with a blue and orange cross.
	
	To understand the rationale behind the notion of incompatible blocks, which is linked to the dotted rectangles around the error vectors in the figure, we need to derive a simple equation from the three errors:
	
	\begin{equation}
	\begin{aligned}
	\mathbf{k} &= \mathbf{x'} - \mathbf{x}, \\
	&= \mathbf{x'} - f_{\text{IDCT}}(\mathbf{d}),\\
	&= \mathbf{x'} - f_{\text{IDCT}}(\mathbf{(c-u) \cdot Q}), \\
	&= \mathbf{x'} - f_{\text{IDCT}}(\mathbf{c \cdot Q}) + f_{\text{IDCT}}(\mathbf{u \cdot Q}),\\
	&= \mathbf{x'} - \mathbf{y} + f_{\text{IDCT}}(\mathbf{u \cdot Q}),\\
	&= \mathbf{e} + f_{\text{IDCT}}(\mathbf{u \cdot Q}).
	\end{aligned}
	\label{eq:errors_link}
	\end{equation}
	The variable $\mathbf{k}$ is the compression error which is the difference between the original pixel block and the decompressed one. If we know the decompressed block and the compression error we can reconstruct an antecedent $\mathbf{\tilde{x}}$ of $\mathbf{c}$. Therefore, if this compression error exists, it directly means that we have an antecedent and by definition the block is compatible. However, this compression error needs to respect a constraint, it has to be an integer by definition.
	
	The other important variable, $\mathbf{u}$, is the rounding error in the DCT domain which is unknown and has to be a floating point value in $\left[-0.5, 0.5\right]$. This constraint defines the dotted rectangles around each spatial rounding error. Finally, we end up with two constraints:
	\begin{enumerate}
		\item the compression error $\mathbf{k}$ needs to be composed of integer values,
		\item $\mathbf{k}$ has to be inside the dotted rectangle to respect the constraint on $\mathbf{u}$.
	\end{enumerate}
	
	We can plot the spatial rounding $\mathbf{e}$ error as shown in Fig.~\ref{fig:example} with a blue cross. The orange cross corresponds to the spatial rounding error of the same block with modifications to its DCT coefficients.
	
	Note that the size of this rectangle is influenced by the quantization matrix $\mathbf{Q}$. The bigger the quantization steps, the larger the rectangle. Consequently, we can speculate that the larger the quantization step, the larger the probability of staying compatible after embedding, which means there exists a quality factor below which every block is compatible. In fact, our empirical experiments show that all blocks are compatible at QF99 and lower so this attack will probably work only at QF100.
	
	\subsection{Finding compatible blocks: the heuristic approach}
	\label{sec:heuristic}
	
	In order to prove that a block is compatible we need to solve the inverse problem \eqref{eq:inverse_prob} where $\mathbf{x}$ is the unknown variable. The variable $\mathbf{c}$ is the observed block (cover or stego) and we assume that both the forward DCT function $f_{\text{DCT}}$ and the quantization table $\mathbf{Q}$ are known.
	A natural potential solution to this problem is the decompressed image $\mathbf{x'} = \left[f_{\text{IDCT}}(\mathbf{Q \times c})\right]$, however most of the time this candidate is not a solution of the problem. For example, in the Alaska dataset~\cite{cogranne2019alaska} composed of 80k images of 1024 blocks, this trivial candidate is a solution in less than 0.5~\% of the blocks with the \texttt{islow} compressor of libjpeg and a naive decompressor.
	
	What is proposed instead, is an algorithm that starts from this candidate and searches for a solution around it by applying small modifications at every pixel position in the block. It behaves like a tree search where each candidate can be extended into new candidates until a solution is found.
	
	For JPEG images compressed at QF100, we can observe that the original pixel block differs from the starting point $\mathbf{x}'$ by only a few modifications of $\pm1$. This is not the case for only 1.7~\textperthousand{} of blocks in the Alaska dataset but it is the case for every block if we ignore the blocks that have been clipped (the decompressed value is below 0 or bigger than 255). This means that if the image is compatible, we should find a solution by potentially applying only $\pm1$ on very specific pixel positions. Moreover, it also gives us an upper bound on the number of candidates: $3^{64}$, which is too large for a brute-force approach.

	Instead of brute forcing every solution, we use a metric to orientate the search toward "good" candidates. For every candidate $\mathbf{\tilde{x}}$, the metric $g$ is the following:
	\begin{equation}\label{eq:metric}
	g_{\mathbf{c}}(\mathbf{\tilde{x}}) = \left\Vert \mathbf{c} - \frac{f_{\text{DCT}}(\mathbf{\tilde{x}})}{\mathbf{Q}} \right\Vert_{\infty},
	\end{equation}
	It corresponds to the infinity norm between the observed DCT block and the DCT block of a candidate. The infinity norm can be motivated by its geometrical form: the set of vectors whose infinity norm is a constant value forms a hypercube. This is exactly what can be observed in the example in Fig.~\ref{fig:example}. The metric of a candidate is defined as the biggest absolute distance between itself and the DCT block $\mathbf{c}$ in the Euclidean DCT space.
	\begin{algorithm}[t]
		\caption{Heuristic to find pixel antecedents}
		\begin{algorithmic}
			\REQUIRE $\mathbf{c} \in \mathbb{Z}^{64}$ \COMMENT{DCT Target block}
			\REQUIRE $\mathbf{Q} \in \mathbb{Z}^{64}$ \COMMENT{Quantization table}
			\REQUIRE $f$ \COMMENT{Forward DCT function}
			\REQUIRE $M \in \mathbb{N}$ \COMMENT{Max iteration}
			\STATE{}
			\STATE $\mathbf{x}' \gets \left[f(\mathbf{Q \times c})\right]$ \COMMENT{Starting point in integer pixel domain}
			\IF {$g_{\mathbf{c}}(\mathbf{x}') < 1/2$}
			\RETURN $\mathbf{x}'$
			\ELSE
			\STATE add $\mathbf{x}'$ to $P_q$ \COMMENT{$P_q$ is a priority queue}
			\ENDIF
			\STATE{}
			\WHILE{$P_q \text{ not empty}$ \AND $i \leq M$}
			\STATE $i \gets i + 1$
			\STATE $\mathbf{\tilde{x}} \gets \text{pop first element of } P_q$ \COMMENT{$\mathbf{\tilde{x}}$ is a candidate}
			\STATE{}
			\FOR {$\mathbf{\tilde{x}}_n \textbf{ in } neighbours(\mathbf{\tilde{x}})$}
			\IF {$\mathbf{\tilde{x}}_n \text{ has not been visited}$}
			\STATE $v_{n} \gets g_{\mathbf{c}}(\mathbf{\tilde{x}}_n)$ \COMMENT{DCT error of $\mathbf{\tilde{x}}_n$}
			\IF {$v_{n} < 1/2$} 
			\RETURN $\mathbf{\tilde{x}}_n$ \COMMENT{$\mathbf{\tilde{x}}_n$ is an antecedent of $\mathbf{c}$}
			\ELSE
			\STATE {}\COMMENT{$\mathbf{\tilde{x}}_n$ is not an antecedent of $\mathbf{c}$ but we store it to explore its neighbours}
			\STATE add $\mathbf{\tilde{x}}_n$ to $P_q$ with priority value $v_{n}$
			\ENDIF
			\ENDIF
			\ENDFOR
			\ENDWHILE
		\end{algorithmic}
		\label{alg:heuristic}
	\end{algorithm}
	In particular, if this metric is lower than half of the size of the hypercube, the candidate is inside the hypercube. In our case, after quantization, the size of the hypercube is 1 so we can briefly show that $g_{\mathbf{c}}(\mathbf{\tilde{x}}) < \frac{1}{2}$ is enough to prove that a block is compatible:
	\begin{equation}
	\begin{aligned}
	&& g_{\mathbf{c}}(\mathbf{\tilde{x}}) < 0.5, \\
	&\Leftrightarrow \forall \,i, & -0.5 \leq c_i - \frac{f_{\text{DCT}}(\mathbf{\tilde{x}})_i}{Q_i} \leq 0.5,\\
	&\Leftrightarrow \forall \,i, & c_i - 0.5 \leq \frac{f_{\text{DCT}}(\mathbf{\tilde{x}})_i}{Q_i} \leq c_i + 0.5,\\
	&\Leftrightarrow \forall \,i, &\left[\frac{f_{\text{DCT}}(\mathbf{\tilde{x}})_i}{Q_i}\right] = c_i.
	\end{aligned}
	\label{eq:proof}
	\end{equation}
	Therefore $\mathbf{\tilde{x}}$ is solution of the inverse problem~\eqref{eq:inverse_prob} and thus the block $\mathbf{c}$ is compatible.\\
	
	The pseudo-code is written in Algorithm~\ref{alg:heuristic} and has some similarity with path-finding algorithms such as A-star~\cite{hart_formal_1968} where each candidate is stored in a priority queue depending on the metric $g_{\mathbf{c}}$. A priority queue is a data structure similar to a list where each element of that list has a priority value associated to it. The element with the smallest priority value is always the first of the list. To put a new element in the queue, we need to store the priority value of each element already in it. In the algorithm, the $v_n$ stands for the priority value of the element $\mathbf{\tilde{x}}_n$ and the quantity associated with this priority value is the metric $g_{\mathbf{c}}(\mathbf{\tilde{x}}_n)$. Using equation~\eqref{eq:proof}, if the metric is smaller than $1/2$, then it is enough to terminate the search algorithm and return the antecedent.
	
	The main advantage of this algorithm is its flexibility to the JPEG pipeline since it is only a called function that can be changed very easily. Moreover, whenever the JPEG pipeline matches the one used to compress the image, it means that by construction if a solution exists it will be found, and therefore it is impossible to have a compatibility error, {\it i.e.} a false negative. On the other hand, the main drawback of this heuristic is that blocks can be proven to be incompatible only if each candidate has been explored which is the worst-case scenario and has a complexity similar to a brute force search. In other words, it is impossible to prove that a block is incompatible in a reasonable time using this algorithm.
	\subsection{Finding incompatible blocks: the ILP approach}
	\label{sec:ilp}
	The other approach is similar to the one presented in~\cite{levecque:hal-04098582}. We use the properties of the DCT rounding error to derive a simple Integer Linear Programming (ILP) problem that can be solved by any solver. To formulate the problem we recall the equation between errors obtained in equation~\eqref{eq:errors_link}:
	\begin{equation*}
	\mathbf{k} = \mathbf{e} + f_{\text{IDCT}}(\mathbf{u \cdot Q})
	\end{equation*}
	This equation is only valid under a few assumptions, the forward DCT has to be:
	\begin{itemize}
		\item linear,
		\item applicable to floating point values since $\mathbf{u}$ is not integer,
		\item the inverse of the IDCT, {\it i.e.} $f_{\text{IDCT}}(f_{\text{DCT}}(\mathbf{x})) = \mathbf{x}$.
	\end{itemize}
	Those assumptions can sound odd if we think about the mathematical definition of the DCT that checks all three of them, but in reality, lots of different DCT implementations exist and only a few of them can satisfy those assumptions.
	
	Using equation \eqref{eq:errors_link} again, we rewrite it to have a formula for the DCT rounding error:
	\begin{equation}
	\mathbf{u} = \frac{f_{\text{DCT}}(\mathbf{k - e})}{\mathbf{Q}}.
	\end{equation}
	This relation is interesting because it links the compression error $\mathbf{k}$ with the DCT rounding error $\mathbf{u}$. Both are unknown variables in our inverse problem but they have two properties:
	\begin{enumerate}
		\item The compression error has to be integer-valued
		\item The DCT rounding error has to be bounded inside a unit cube.
	\end{enumerate}
	
	Therefore we have the following problem:
	\begin{alignat*}{2}
	\text{find } & \qquad \mathbf{k} \in \mathbb{Z}^{64}, &\\
	\text{such that } & \qquad \left|\mathbf{u}\right| = \left|f_{\text{DCT}}(\mathbf{k - e})\right| \leq \frac{1}{2} &\text{ element-wise.}
	\end{alignat*}
	This problem can be solved by any ILP solver, in our case, we used \texttt{Gurobi} through the python \texttt{Gurobipy} module. The worst-case scenario is also reached when the problem is infeasible which means that the block is incompatible. However, the main advantage compared to the heuristic approach is that such a solver can prove in a realistic time that a block is incompatible. The main drawback of this approach is that it will only work with a JPEG pipeline using a mathematical definition of the DCT, {\it i.e.} the naive DCT.
	
	To sum up this section, we have seen that the compatibility problem can be formulated as the inverse problem~\eqref{eq:inverse_prob}. To solve this problem there are two approaches proposed in this paper, the heuristic one is fast for compatible blocks and can deal with every DCT pipeline but it cannot prove that a block is incompatible. The ILP approach is slower and only works with the naive DCT but it can prove that a block is incompatible after a certain amount of time (at least 10h for a single block with \texttt{Gurobi} on an Intel Xeon at 2,2 GHz and 8 threads).
	\section{Statistics on incompatible blocks}
	\label{sec:statistics}
	This section presents two important statistics for the rest of the paper: the probability for a block to become incompatible, and the connections between incompatibility and other features, such as the variance of the spatial rounding error and the position of the modified DCT coefficient inside the block.
	\subsection{Probability of becoming incompatible}\label{sec:proba}
	\begin{figure}[t]
		\centering
		\includegraphics[width=\linewidth]{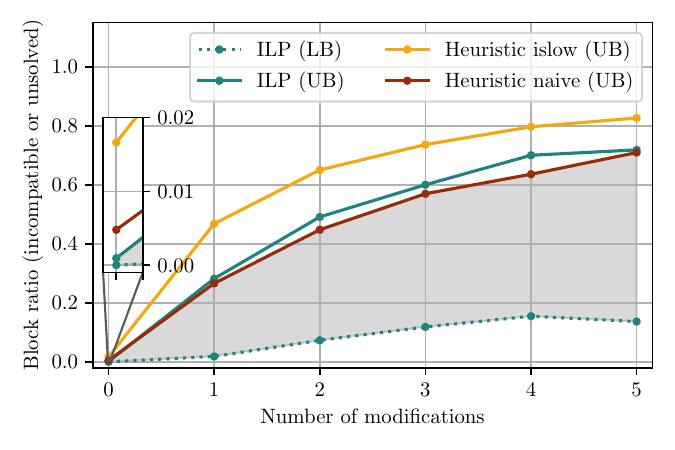}
		\vspace{-3em}
		\caption{Empirical probability for a block to become incompatible after a given number of $\pm1$ modifications on DCT coefficients. The dashed cyan curve corresponds to blocks proven incompatible using the ILP approach after 3B iterations, it is a lower bound (LB) of the real value. The plain cyan curve shows incompatible and unsolved blocks using the ILP approach again, it is an upper bound (UB) of the real value. Finally, the red and yellow curves show unsolved blocks: another upper bound for 50k iterations of the heuristic approach for two different compressors. Cyan used 100 blocks per point and red and yellow used 1000 blocks per point.
		}
		\label{fig:proba}
	\end{figure}
	Using the Alaska dataset, we designed an experiment to track how many blocks can become incompatible depending on the number of modifications. It will give an approximation of the probability for a block to become incompatible after a given number of modifications. This statistic is not the main result of the paper but it will be important to build our steganalysis detector.
	The experiment is built as follows: we extract a random block $\mathbf{x}$ from a random image from Alaska that will be compressed using the naive and \texttt{islow} methods to obtain the DCT block $\mathbf{c}$. Depending on the number of modifications $m$ we select $m$ coefficients without replacement and randomly add $\pm1$. This block is then sent to the two algorithms (ILP and heuristic approach) to try to find an antecedent and try to prove that it is still compatible. The heuristic algorithm runs for 50k iterations (a few minutes in the worst case scenario) and the ILP algorithm runs for 3B iterations of \texttt{Gurobi} (approximately 20h in the worst case scenario). There are 3 possible results:
	\begin{itemize}
		\item if an antecedent is found, the block is compatible,
		\item if no antecedent is found and the algorithm reaches the time-out value, the block is unsolved,
		\item for the ILP approach, the solver can prove that a block is incompatible because no antecedent exists.
	\end{itemize}
	In order to understand the future results, we need to introduce some notations. Let $H$ and $\text{ILP}$ be used as indexes to identify variables associated with the heuristic algorithm and the ILP approach. Let $C_H$ and $C_{\text{ILP}}$ be the empirical ratios of proved compatible blocks, $I_H$ and $I_{\text{ILP}}$ the empirical ratios of proved incompatible blocks, and $U_H$ and $U_{\text{ILP}}$ the empirical ratios of unsolved blocks. In those last ratios, some of them are compatible $\tilde{C}_H$, $\tilde{C}_{\text{ILP}}$ and some of them are incompatible $\tilde{I}_H$, $\tilde{I}_{\text{ILP}}$ but with unknown values such that: $U_H = \tilde{C}_H + \tilde{I}_H$ and $U_{\text{ILP}} = \tilde{C}_{\text{ILP}} + \tilde{I}_{\text{ILP}}$. All those values depend on the number of iterations, but for the sake of clarity, this value is omitted here. Finally, the true probability to be compatible is approximated by $p_C = C_H + \tilde{C}_H = C_{\text{ILP}} + \tilde{C}_{\text{ILP}}$ and the true probability to be incompatible is approximated by $p_I = I_H + \tilde{I}_H = I_{\text{ILP}} + \tilde{I}_{\text{ILP}}$.
	
	For the heuristic algorithm, we use 1k blocks for each number of modification 
	but 100k blocks if there is no modification since all blocks are compatible and the algorithm is very fast in that case. However, there is no incompatible block detected with this approach: $I_H = 0$, so we approximate the probability of incompatible blocks by the ratio of unsolved blocks $U_H$. Because of the time-out value, we can show that this approximation is an upper bound (UB): $U_H =  \tilde{C}_H + \tilde{I}_H \geq \tilde{I}_H = p_I$. But this approximation is only good if the ratio of unsolved compatible blocks $\tilde{C}_H$ is small.
	
	For the ILP approach, we use 100 blocks per point (the method is very slow) and we can extract a lower bound and an upper bound of the probability of being incompatible. First, $I_{\text{ILP}}$ is an lower bound (LB): $I_{\text{ILP}} \leq p_I$. Second, the ratio of incompatible plus the ratio of unsolved blocks is an upper bound of the probability to be incompatible: $I_{\text{ILP}} + U_{\text{ILP}} = I_{\text{ILP}} + \tilde{C}_{\text{ILP}} + \tilde{I}_{\text{ILP}} \geq I_{\text{ILP}} + \tilde{I}_{\text{ILP}} = p_I$.
	
	\begin{figure}[t]
		\centering
		\includegraphics[width=\linewidth]{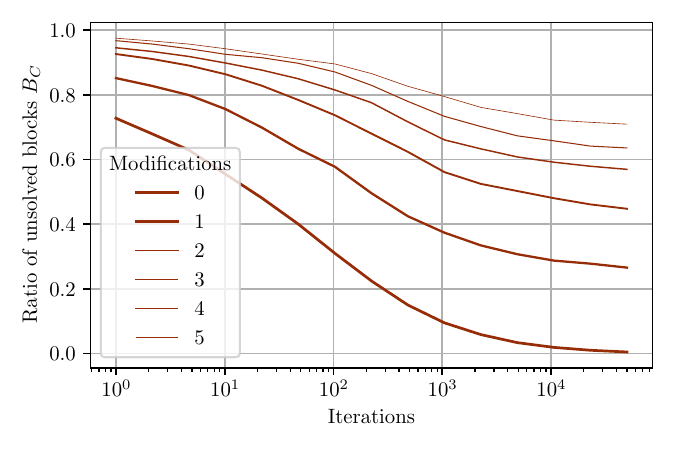}
		\vspace{-3em}
		\caption{Variation of unsolved blocks using the heuristic algorithm depending on the number of iterations and modifications. Convergence slows down after 50k iterations, which suggests that the heuristic upper bound is not very far from the real value.}
		\label{fig:proba_variation}
	\end{figure}
	
	Finally, we have two upper bounds and one lower bound: $p_I \leq U_H$ for the heuristic algorithm and $I_{\text{ILP}} \leq p_I \leq I_{\text{ILP}} + U_{\text{ILP}}$ for the ILP algorithm.
	
	Results of this experiment are shown in Fig.~\ref{fig:proba} and we can conclude that the true probability $p_I$ of becoming incompatible after a number of modifications is between the lower bound of the ILP approach and the minimum of the two upper bounds. The other interesting result is that for 0 modifications, the ratio of unsolved blocks $\tilde{B}$ of the heuristic algorithm is approximately $0.5\%$. That gives an idea of the gap between the upper bound of the heuristic and the true probability (which is exactly 0 when there is no modification): the true value is probably closer (give or take a few percent) to the upper bound than to the lower bound.
	
	The yellow line shows the upper bound of the heuristic algorithm but using another DCT method which is \texttt{islow}. We can see that this upper bound is always above the one from the naive DCT method even for 0 modification by a significant gap. We propose two interpretations: either the probability of being incompatible is very dependent on the DCT method used, or the heuristic algorithm is not as good at solving the problem with this DCT method. The second interpretation seems more meaningful when we dive deeper into the heuristic algorithm implementation. See Appendix~\ref{sec:appendix_heuristic_metric} for more details.
	
	Fig.~\ref{fig:proba_variation} shows the evolution of the ratios of unsolved blocks $U_H$ using the heuristic algorithm depending on the number of iterations. At 50k iterations, those values - and therefore the upper bound on the probability of incompatible - start to stagnate and we can see that with a few iterations more, all 100k blocks with 0 modifications could have been solved.
	\subsection{Links with other statistics}
	\begin{figure}[t]
		\centering
		\includegraphics[width=0.70\linewidth]{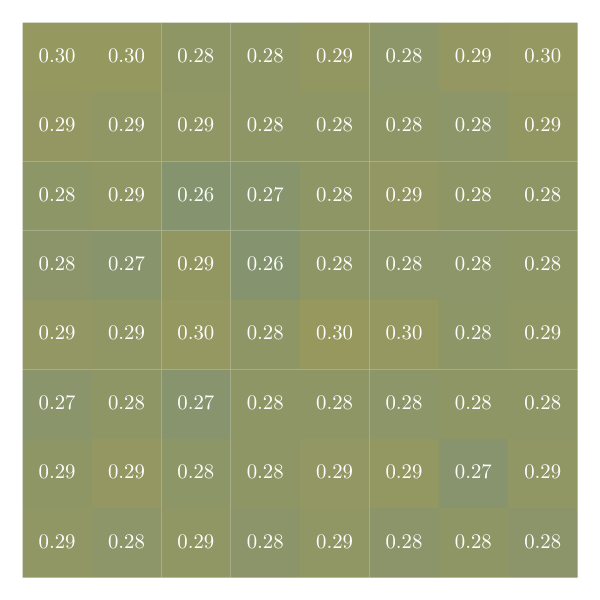}
		\vspace{-1em}
		\caption{Heat-map showing the probability for a block to become incompatible after modifying a single coefficient for every one of the 64 positions. The randomness of the pattern tells us that there is no dependence between the position of the modification and the incompatibility of the block.}
		\label{fig:heatmap}
	\end{figure}
	The previous paragraph shows a clear correlation between the number of modifications and the probability of becoming incompatible. A natural question that emerges from this result is whether the incompatibility depends on the position of the modification. Is there more chance of becoming incompatible if we modify the DC or an AC coefficient? We ran a similar experiment as before but this time the modifications were not randomly selected. For every block, we try to find an antecedent after applying a single modification in each of the 64 coefficients for $+1$ and $-1$ separately. Using the heuristic algorithm we cannot prove that a block is incompatible, but we can use the same approximation as before with the ratio of unsolved blocks $U_H$. The heat map in Fig.~\ref{fig:heatmap} shows that this probability is uniform for every position (also the case if we separate the $+1$ for the $-1$). We conclude from these results that the incompatibility does not depend on the position of the change or the sign of the change. This can be explained by the fact that the DCT transform can be seen as an arbitrary rotation of the block in the pixel domain to the block in the DCT domain and that there is no reason that one direction is more subject to incompatibility than another. This property will be used in the next section to build a statistical test.
	
\begin{figure}[t]
		\centering
		\includegraphics[width=\linewidth]{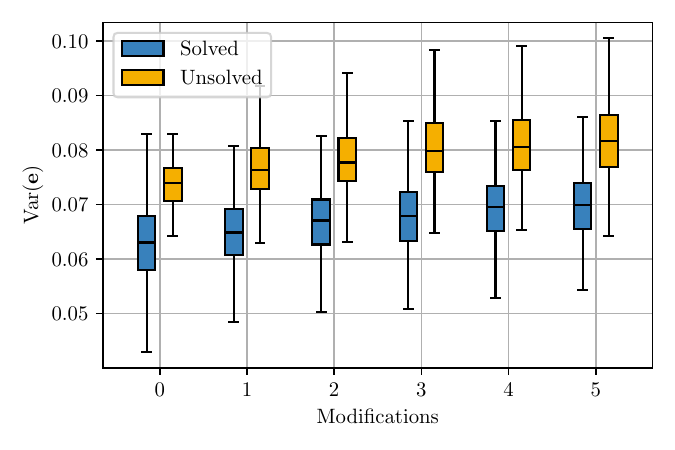}
		\vspace{-3em}
		\caption{Correlation between the number of modifications and the variance of the spatial rounding error. Outliers are hidden for the sake of clarity. There is a clear correlation between a high variance and a high probability of being incompatible but also a correlation between the variance and the number of modifications.}
		\label{fig:variance}
	\end{figure}

	The second feature is the variance of the spatial rounding error. This feature is the cornerstone of the RJCA~\cite{butora2019reverse} because we can show that at high QF, the distribution under the cover hypothesis converges toward a wrapped Gaussian with a known variance close to $0.064$. But under the stego hypothesis, the same variance increases toward $\frac{1}{12}\simeq 0.083$, the variance of a uniform distribution. In Fig.~\ref{fig:variance} we observe a very strong correlation between the variance of a single block and the compatibility property. In particular, the variance of the spatial rounding error (a very easy feature to compute) could be used as a proxy for compatibility and also for the number of modifications. Indeed, we can see that for unsolved blocks, the more coefficients are modified, the higher the variance.
	
	\section{Steganalysis}\label{sec:steganalysis}
	
	In the framework of steganography and steganalysis, by definition, an incompatible block is a block that has been modified. Previous sections have presented algorithms to track the compatibility or not of a block and some statistics about the probability of existence. In this section, we propose two scenarios of steganalysis: the first one uses the ILP approach assuming a lot of resources (time especially) and the second one is a Likelihood Ratio Test using the heuristic approach.
	
	\subsection{Zero False Alarm steganalysis} \label{sec:zero_false_alarm_steganalysis}
	
	The main advantage of the ILP approach for finding incompatible blocks is precision: if the solver proves that the ILP problem is infeasible, then the block does not have any antecedent and has been modified by the embedding: this deduction is always true. Therefore, the ILP method gives a perfectly precise classification and does not produce any False Alarm (misclassifying a cover image). However, this precision has an expensive price which is the amount of resources needed in time and processing power.
	
	We will assume that we have infinite resources to carry out our experiments in the form of a simulation since as seen in Section~\ref{sec:ilp}, the detection of a single incompatible block is already very time-consuming. On the Alaska dataset, we embed images of $N = 1024$ blocks using J-UNIWARD and UERD for different embedding rates. We count the number of modifications $m_i$ for every block $i=1,\ldots,N$, and we use the probabilities $p_m$ to become incompatible with the ILP solver obtained in Fig.~\ref{fig:proba}.
	
	The probability $P_{\text{incomp}}$ that at least one block of the image is detected as incompatible can be derived as follows:
	\begin{equation}
	P_{\text{incomp}} = 1 - \prod_{i = 1}^{N} \left(1 - p_{m_i}\right).
	\end{equation}
	For example, using two blocks and using the empirical results of section~\ref{sec:proba}, the first with three modifications and the second with one modification, the probability of being detected for at least one block (and thus for the image) is $1 - (1 - p_3)(1 - p_1) = 1 - 0.85\times0.97 \approx 0.18$.
	
\begin{figure}[t]
		\centering
		\includegraphics[width=0.9\linewidth]{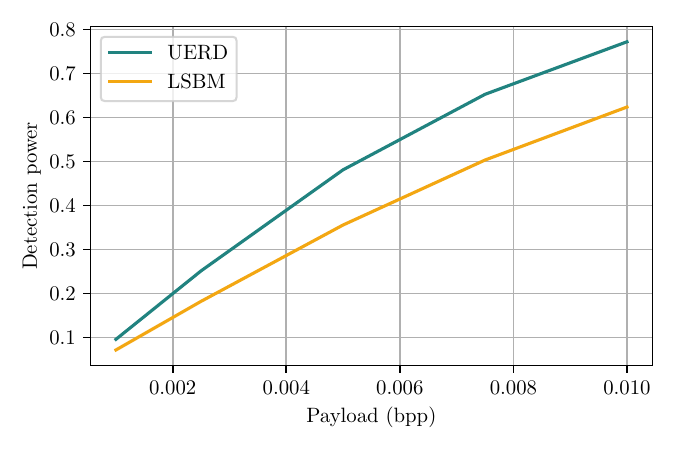}
		\vspace{-1.5em}
		\caption{Detection power depending on the embedding rate with the ILP approach for two popular embedding schemes (UERD and LSBM) assuming infinite resources for an image of 1024 blocks. An image is detected as stego if at least one block is incompatible. Note that this detection method does not generate false positives.}
		\label{fig:zero_false_alarm}
	\end{figure}

	Results of this simulation are shown in Fig.~\ref{fig:zero_false_alarm}, where we can see that the detection increases with the payload and tends toward perfect detection around 0.03 bpp. Notice how UERD is more detectable than LSBM because it modifies more coefficients and thus it is more susceptible to creating incompatible blocks. It should also be remembered that the probabilities used for this simulation are a lower bound of the true probabilities and hence the true detection power should be greater than the one in the graph if more iterations of the algorithm were used.
	
	\subsection{Likelihood Ratio Test} \label{sec:likelihood_ratio_test}
	
	Proving that a block is compatible is easier than the opposite but it is not sufficient to decide if the image is cover or not.
	
	Instead, we now derive a Likelihood Ratio Test (LRT) on the amount of unsolved blocks in an image. Unsolved blocks are the ones for which there is still no antecedent found after running a given number of iterations of an algorithm. We propose in this subsection to build a statistical test based on the number of unsolvable blocks found in a test image. Since the heuristic algorithm is faster and more versatile, we decided to use it to detect unsolvable blocks.
	
	In order to build our test we introduce a few variables. Let $i \in \{1,\dots,n\}$ be the indices of blocks with $n\leq N$ an hyper-parameter. We use the heuristic algorithm on every block, but since this algorithm cannot prove that a block is incompatible (because of the complexity of the problem) we observe a vector of variables $\mathbf{T} = (t_1, \dots, t_n)$. If $t_i = 1$ the algorithm reached the timeout value for the block $i$ and did not find any antecedent, it is consequently unsolvable for a given amount of iterations. Otherwise, if $t_i=0$ the block is compatible. We have seen in the last section that the probability for a block to become incompatible depends only on the number of modifications $m_i$ in this block. So, let $\mathbf{M} = (m_1, \dots, m_n)$ be the vector of number of modifications. We define the likelihood for an image to observe a vector $\mathbf{T}$ as $p_{\mathbf{M}}(\mathbf{T}) = p_{m_1}(t_1) \times \dots \times p_{m_n}(t_n)$. 
	
	We consider two hypotheses:
	\begin{equation}
	\begin{aligned}
	\mathcal{H}_0&: \text{The image is cover:}\quad \mathbf{T} \sim p_\mathbf{0}, \\
	\mathcal{H}_1&: \text{The image is stego:}\quad \mathbf{T} \sim p_{\mathbf{M}}, \,\forall\, i,\, m_i\geq 0.
	\end{aligned}
	\end{equation}
	Because of the independence between blocks, the likelihood ratio for the image is the product of the likelihood ratio of every block:
	\begin{equation}
	\begin{aligned}
	\Lambda(T) &= \frac{p_{\mathbf{M}}(\mathbf{T})}{p_0(\mathbf{T})}\\
	&= \prod_{i = 1}^{n} \frac{p_{m_i}(t_i)}{p_0(t_i)}\\
	&= \prod_{i = 1}^{n}  \frac{\sum_{m = 0}^{64} P(t_i |m_i = m) P(m_i = m)}{P(t_i | m_i = 0)}
	\end{aligned}
	\end{equation}
	Finally, the log-likelihood ratio for an image using $n$ blocks is:
	\begin{equation}\label{eq:LRT}
	\begin{aligned}
	\log(\Lambda(\mathbf{T})) = \sum_{i = 1}^{n} \Biggl[&\log\left(\sum_{m = 0}^{64} P(t_i |m_i = m) P(m_i = m)\right) \\
	&- \log(P(t_i | m_i = 0)) \Biggr]
	\end{aligned}
	\end{equation}
	
	We now give the analytic form of the prior $P(m)$ and the values of likelihood $P(t|m)$.
	
	\textbf{Prior.} The prior on the number of modifications $P(m)$ can be crucial to yield a powerful test. Therefore, we make a separation of cases:
	
	The easiest case and probably the least powerful is when nothing is known. In this case, the prior on $m$ is a uniform law on $\left[1;64\right]$.
	
	Another alternative happens when the detector is Selection-Channel-Aware~\cite{denemark2014selection}: if the map of embedding probabilities (aka the p-map) of the embedding algorithm is known we can use it to derive the approximate number of modifications. The p-map of a block is composed of 64 probabilities $(q_i)_{i \in \left[1;64\right]}$, one for each coefficient, and represents the probability to be modified. The distribution of the number of modifications is given by the Poisson-Binomial law which is the sum of 64 Bernoulli experiments all with different parameters. We denote this law $\mathcal{P_B}((q_i)_{i \in \left[1;64\right]})$.\\
	
	To sum up the different priors we use are:
	\begin{itemize}
		\item nothing is known: we then use the classical uniform prior, {\it i.e.} $P(m) \sim \mathcal{U}(\left[1;64\right])$,
		\item the embedding algorithm and the p-map are known: $P(m) \sim \mathcal{P_B}((q_i)_{i \in \left[1;64\right]})$.\\
	\end{itemize}
	
	The likelihoods $p_m(t) = P(t|m)$ can be derived from Fig.~\ref{fig:proba}. Note that the likelihood and $t$ depend on a hidden parameter which is the maximum number of iterations. For the sake of simplicity, we do not write the number of iterations explicitly but it is important to take it into account as a parameter that will impact the LRT, hence the classification performances.
	
	\subsection{Selection strategies} \label{sec:selection_strategies}
	The LRT designed in the previous paragraph can scale with the number of blocks available in the image. However, the algorithms solving the inverse problem are costly in time. We are thus facing a trade-off between resources and performance: the image is composed of $N$ blocks but we only have the resources to solve the problem on $n < N$ blocks, how do we choose them to maximize the detection performance?
	
	We present three different strategies to sort blocks of an image so that we increase our chance of selecting blocks susceptible to being modified and thus potentially incompatible.
	
	\begin{itemize}
		\item \textbf{Random:} The random strategy is the trivial one: the blocks are randomly sorted. It will be used to compare with other strategies.
		\item \textbf{Variance of the rounding error:} The RJCA~\cite{butora2019reverse} was built on the study of the variance of the spatial rounding error. In the cover image, this variance should be fixed but under the stego hypothesis, this variance increases. It seems reasonable to think that the same behavior happens for blocks: modified blocks should be more likely to have a bigger variance of the spatial rounding error than cover blocks. This behavior was already observed~\cite{butora2022fighting} for different embedding algorithms and rounding functions.
		\item \textbf{SCA:} Assuming selection-channel-awareness means that we know the probability of modifications for every DCT coefficient. In this case, it is possible to sort blocks depending on the expected modification number. This value is simply the mean of the 64 probabilities of the probability map in a block.
	\end{itemize}
	\begin{figure*}[t]
		\centering
		\resizebox{1\textwidth}{!}
		{
			\def\nodedistance{1}
			\def\xsep{1.4}
			\def\ysep{0.9}
			\tikzstyle{box} = [rectangle, text centered]
			\tikzstyle{arrow} = [thick,->,>=stealth]
			\tikzstyle{textbox} = [rectangle, align=center, draw, minimum height=1* \ysep cm,minimum width=2* \xsep cm]
			\tikzstyle{coloredtextbox} = [rectangle, align=center, draw=black, fill={rgb, 255:red, 249; green, 231; blue, 202}, draw={rgb, 255:red, 242; green, 202; blue, 140}, thick, minimum height=1* \ysep cm,minimum width=2* \xsep cm]
			\tikzstyle{textellipse} = [ellipse, align=center, draw=black, thick, minimum height=1.5* \ysep cm, minimum width=2.5* \xsep cm, fill=white]
			\begin{tikzpicture}[node distance=\nodedistance cm]
			\node[textbox, draw=red] (image) {JPEG Image};
			\node[textellipse, right=of image] (selection) {\textbf{Block selection strategy}~\ref{sec:selection_strategies}\\(SCA, Variance or Random)};
			\node[textbox, right=of selection] (blocks) {Block subset};
			\node[textellipse, right=of blocks] (algorithm) {\textbf{Algorithm~\ref{alg:heuristic}}\\on selected blocks};
			\node[textellipse, draw=red, below=of algorithm] (pipeline) {JPEG pipeline\\of the image};
			\node[textellipse, right=of algorithm] (lrt) {\textbf{LRT}~\ref{eq:LRT}};
			\node[textbox] (likelihood) at (lrt|- pipeline) {Likelihood\\(see Fig.~\ref{fig:proba})};
			\node[textbox, above=of lrt, draw=red, dashed] (prior) {Change rate prior\\(SCA or Uniform)};

			\draw[arrow] (image.east) -- (selection.west);
			\draw[arrow] (selection.east) -- (blocks.west);
			\draw[arrow] (blocks.east) -- (algorithm.west);
			\draw[arrow] (algorithm.east) -- (lrt.west);
			\draw[arrow] (pipeline.east) -- (likelihood.west);
			\draw[arrow] (pipeline.north) -- (algorithm.south);
			\draw[arrow] (likelihood.north) -- (lrt.south);
			\draw[arrow] (prior.south) -- (lrt.north);
			\end{tikzpicture}
		}
		\caption{Flowchart for the steganalysis detector based on the heuristic algorithm and the LRT. First a selection strategy is used to select a subset of the blocks, then the heuristic algorithm is applied on all blocks and the LRT is computed using the results of the algorithm and the likelihood function computed knowing the pipeline. The red color is used to identify what is known by the agent performing the analysis.}
		\label{fig:flowchart}
	\end{figure*}
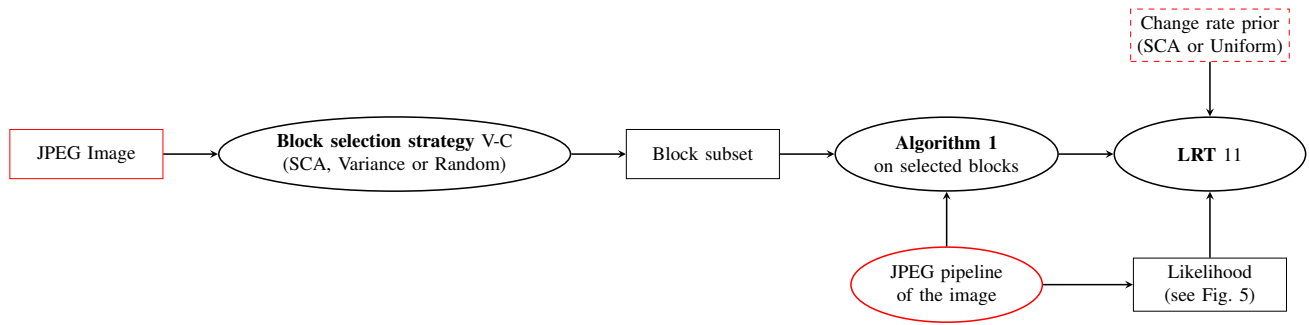

	\section{Results}\label{sec:results}
	\begin{table}[h]
		\centering
		\renewcommand\arraystretch{1.5}
		\normalsize
		\begin{tabular}{c|c|c}
			\textbf{Name} & \textbf{Block selection} & \textbf{Prior} \\ \hline
			\textit{SCA}           & SCA                      & SCA            \\ \hline
			\textit{Blind}         & Variance                 & Uniform        \\ \hline
			\textit{Partial SCA}   & SCA                      & Uniform        \\ \hline
			\textit{Control}       & Random                   & Uniform
		\end{tabular}
		\caption{Combination of block selection strategies with priors.}
		\label{table:notations}
	\end{table}
	\subsection{Experimental setup}

	In order to show the main performances of the LRT for small datasets of images we will use a simulation instead of a real run which would take a lot of time and resources. This simulation is possible because we assume that incompatibility only depends on the number of modifications in a DCT block (\textit{cf}. Fig.~\ref{fig:proba}). In particular, the property does not depend on modification positions (\textit{cf}. Fig.~\ref{fig:heatmap}) or the content which means that the embedding scheme does not matter, only the number of modifications matters. We show in one experiment (see Fig.~\ref{fig:roc_esrnet}) that the simulation is meaningful and very close to classification with image databases.
	
	We used 5000 images of size $256 \times 256$ from Alaska and embedded them using UERD and LSBM at different payloads between 0.001 and 0.01 bits per pixel (bpp). Instead of running the heuristic on every block, we simulate the output as follows. Let $\mathbf{M} = (m_0, \dots, m_N)$ be the vector of the number of modifications per block for an image composed of $N$ blocks. We want to simulate the output of the heuristic algorithm $\mathbf{T} = (t_0, \dots, t_N)$ to compute the LRT test. To do so, we draw a random vector $\mathbf{r} \in [0;1]^N$ and by comparing it with the probabilities to be incompatible knowing the number of modifications, we can simulate the output of the algorithm:
	\begin{equation}
	\forall\, i, \quad t_i = \left\{
	\begin{aligned}
	1 &\text{ if } r_i \leq P(t=1 | m_i),\\
	0 &\text{ otherwise}.
	\end{aligned}
	\right.
	\end{equation}
	Where the probabilities $P(t=1|m)$ are given in Fig.\ref{fig:proba}.
	
	With this simulation of $\mathbf{T}$ we can compute the LR with formula~\eqref{eq:LRT}. The vector $\mathbf{T}$ can be sub-sampled with a selection strategy of blocks that are more susceptible to being incompatible (\textit{cf}. section~\ref{sec:selection_strategies}) and the prior of the LR can be adjusted depending the SCA assumption or not.
	
	To show that our simulation is close to reality, we also run the test on a a subset of Alaska dataset composed of only 600 cover images and 600 stego images embedded at 0.01 bpp with LSBM and UERD. Moreover, we also test it on the a subset of BossBase~\cite{bas2011break} dataset composed of 600 cover images and 600 stego images embedded at 0.01 bpp with UERD.
	
	For comparison w.r.t. deep learning detectors, we selected e-SRNet, the SRNet~\cite{srnet} trained on the decompression errors $\mathbf{e}$. We split the 25k $256\times 256$ Alaska images into training, validation, and testing sets of sizes $19$k, $1$k, and $5$k respectively. We trained the detector on $0.01$ bpp for $30$ epochs with the rest of the hyperparameters as described in~\cite{but23ih}. The best checkpoint was used to refine the classifier on smaller payloads for additional $15$ epochs. To ensure a fair comparison, all images in the training set were compressed using the same compressor as the one utilized in our heuristic approach.
	
	\begin{figure}[t]
	\centering
	\includegraphics[width=\linewidth]{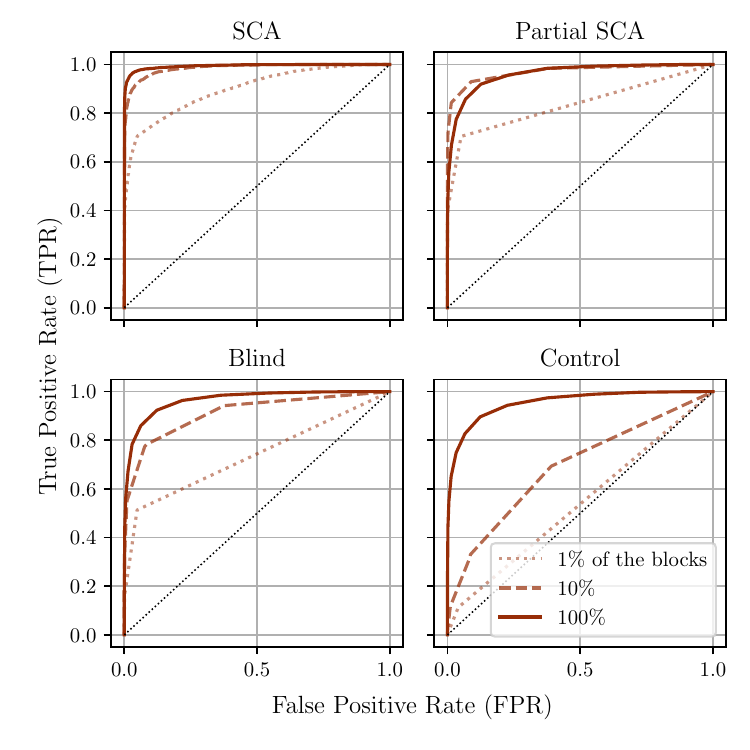}
	\vspace{-3em}
	\caption{Simulation of ROC for different block selection strategies and prior. Stego images are embedded with UERD at 0.005 bpp. The first term of the title is the selection strategy and the second is the prior. }
	\label{fig:roc_strat_uerd}
	\end{figure}

	We presented two different priors and three selection strategies in sections~\ref{sec:likelihood_ratio_test} and~\ref{sec:selection_strategies}, but not every combination is interesting. Assuming selection channel awareness means that the embedding algorithm used and the message length are known, we can therefore compute the probabilities of modifications for every DCT coefficient. The exact prior should be computed with the probabilities of modifications of the cover image. But if this one is not known, we can use the potential stego image to yield a very close approximation of it. On the contrary, if no knowledge is available to perform the steganalysis we can only access the variance of the spatial rounding error. We present the results for four strategies, one with SCA, one without SCA, and two "control" strategies in table~\ref{table:notations}.
	
	Finally, we introduce  the total probability of error under equal priors that will be used to compare different models and strategies:
	\begin{equation*}
	P_E = \min_{P_{\text{FA}}} \frac{P_{\text{FA}} + P_{\text{MD}}(P_{\text{FA}})}{2},
	\end{equation*}
	with $P_{\text{FA}}$ the probability of False Alarm (type I error) and $P_{\text{MD}}$ the probability of Missed Detection (type II error).
	
	\subsection{Results}
	\label{subsec:results}
	\begin{figure}[t]
		\centering
		\includegraphics[width=0.8\linewidth]{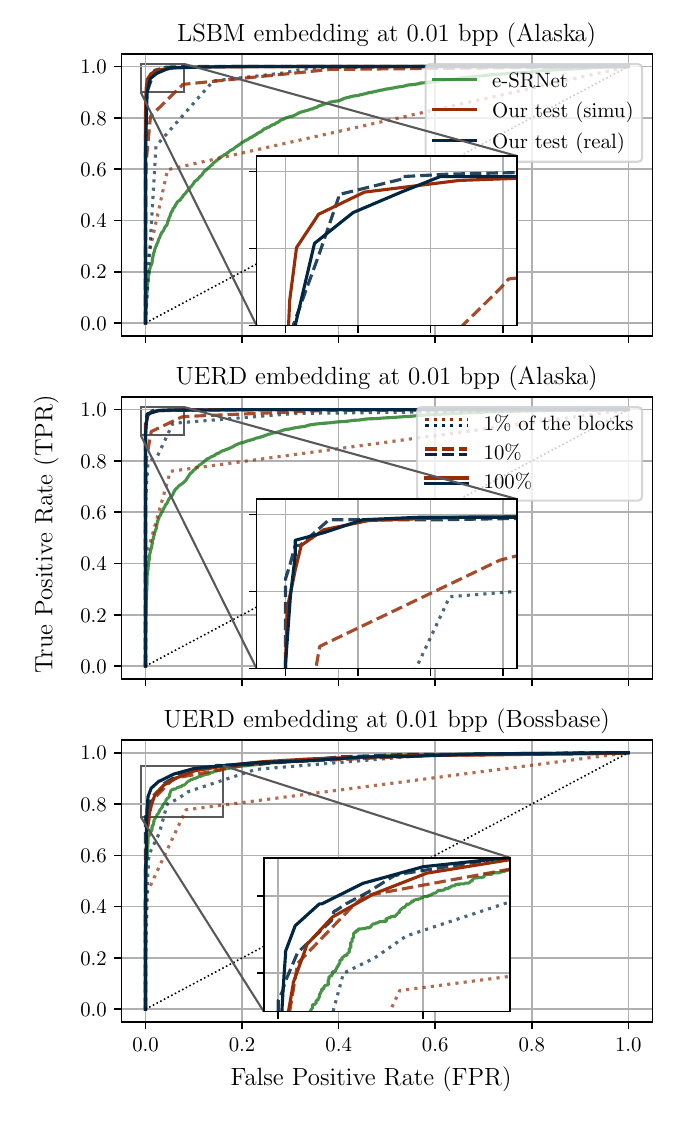}
		\vspace{-1.5em}
		\caption{Comparison of ROCs with e-SRNet for LSBM and UERD at 0.01 bpp. Red lines are simulated and blue ones are a complete realization of our test without simulation. In our results, blocks are selected with the variance of spatial rounding error and the prior used is uniform. We can see the real experiment fits very well the simulation or even outperforms it with 1 and 10\% of the blocks.}
		\label{fig:roc_esrnet}
	\end{figure}
	\begin{figure}[t]
		\centering
		\includegraphics[width=0.8\linewidth]{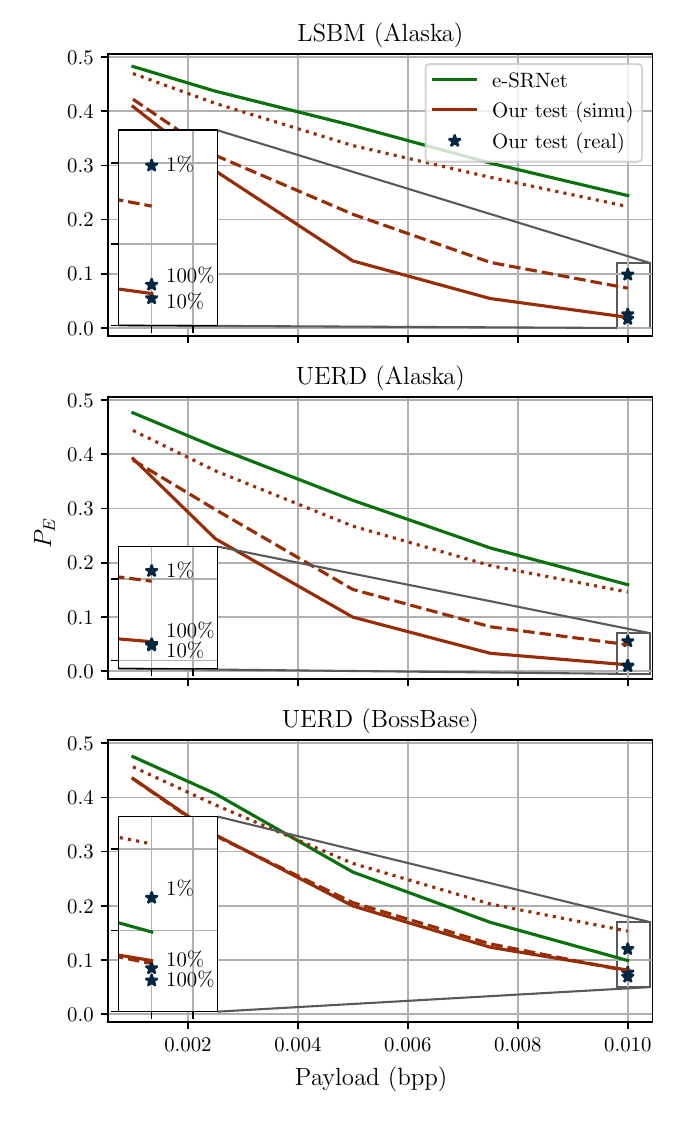}
		\vspace{-1.5em}
		\caption{$P_E$ for the blind strategy and comparison with e-SRNet.}
		\label{fig:pe}
	\end{figure}
	Fig.~\ref{fig:roc_strat_uerd} shows the impact of each strategy using images embedded at 0.005 bpp with UERD. Without surprise, with the SCA assumption, the best strategy is to use it to select blocks and to compute the prior in the LRT. But without SCA assumption, in the blind case, the variance of the spatial rounding error seems to be a good metric to select blocks: the bottom left graph shows that using 10\% of the blocks out of 1024 is already close to the best performance using all blocks. Selecting blocks randomly is a bad choice if you do not use every block in the image, you are most likely to select compatible blocks that do not show any evidence of stego message. 
	
	The partial SCA is interesting because it shows that the LRT performs better when using only 10\% of the blocks if they are well chosen instead of taking the full image. The intuition behind these results is that the LRT becomes more confident that the image is stego when there is a majority of unsolved blocks with high prior due to the SCA assumption among the solved blocks. If $100\%$ of the blocks are used, there is a majority of solved blocks because the payload is so small that only a few blocks are modified and therefore, the LRT tends to give more weight to the cover hypothesis even if there is still a non-negligible fraction of unsolved blocks.
	
	The shape of the ROC can also be explained due to the small number of blocks used in the LRT. If only 10 blocks are used, there are only 11 possible outcomes of the heuristic: either 0, 1, ..., 9, or 10 blocks are unsolved. Therefore the LRT using 10 blocks only has 11 possible values and thus 11 maximum points in the ROC curves. Most of them are close to the point (1,1). In the case of the SCA assumption, the ROC is smooth because different priors are used for each image which yields different values for the LRT and thus an infinite possibilities of points on the ROC.
	
	Note that using the LSBM embedding, SCA, partial SCA, and Control returns the same results because the probabilities of modifications are equal for all elements and for all images (the message always has the same length) and thus do not give any information about the locations of modifications. With LSBM only the blind strategy is useful and will be used for the next results.

	Fig.~\ref{fig:roc_esrnet} represents the ROC for 0.01 bpp embedded with LSBM and UERD on Alaska and Bossbase datasets. The performance of e-SRNet corresponds to the green line and red lines are the simulated result on the same dataset. The blue lines are the results of an experiment without simulation but less images. Dotted lines and dashed lines are the results using only 1\% and 10\% of the blocks of the image respectively. The main feature of this figure is that both simulation and real experiments are very close to each other when using the full image. There are few differences when using 1\% or 10\% of the blocks but we recall that this experiment used only 600 stego images and 600 cover images and thus is sensible to the noise. However, one explanation is that for the simulation, the probability of being incompatible for 6 or more modifications per block is supposed equal to the probability with 5 modifications per block, which means that the probabilities of the simulation are underestimated. And, even at 0.01 bpp, we can observe more than 5 modifications in some blocks (especially with the UERD scheme). The figure also shows that our test has better results than e-SRNet even without the SCA assumption (the blind strategy was used). 
	
	On BossBase images, e-SRNet produces better results compared to our approach, which performs worse than on Alaska. When using the same payload, UERD modifies fewer blocks on average in BossBase than in Alaska, which may account for the slight difference. Note that, the simulation on BossBase was conducted using the likelihood computed on Alaska, highlighting the content-independent nature of our method.
	
	Moreover, the agreement between the simulated and non-simulated results is a good argument in favor of our simulation hypotheses. In particular, the fact that the incompatibility property only depends on the number of modifications per block. We recall that for this set of experiments, we decided to use mainly simulations to reduce greatly the computational time, by decreasing it by one order of magnitude, without altering the scientific conclusions.
	
	Fig.~\ref{fig:pe} shows the evolution of the total probability of error under equal prior $P_E$ for the blind strategy and different payloads between 0.001 and 0.01 bpp. We can see that even with 1\% of the blocks the heuristic algorithm outperforms e-SRNet both on UERD and LSBM embedding schemes. By using more blocks the algorithm tends to perfect detection for bigger payloads. Of course, if the SCA assumption is used, the performance could be even better.
	
	\section{Conclusions and perspectives}
	Incompatible JPEG blocks can be created at QF100 with any embedding scheme. This property does not depend on the position of the modifications but only on the number of modifications per block and is very correlated to the variance of the spatial rounding error. The question of differentiating incompatible blocks from compatible ones has been formulated as an inverse problem with or without a solution. Two algorithms were presented to solve this problem: a heuristic algorithm that can use every JPEG pipeline in a black-box setting and can find the solution rather quickly if it exists, and an ILP formulation that can prove that blocks are incompatible with the naive DCT method.
	
	We used the heuristic algorithm combined with a likelihood ratio test to derive a powerful classifier. Using different selection strategies, it can use only a subset of the blocks in the image and yield state-of-the-art results at QF100. It is also possible to improve it with a selection-channel-aware version that is even better than the blind version.
	
	This detector requires knowledge about the JPEG pipeline used to compress stego and cover images and a certain amount of time to solve the inverse problems but the only limit to its performance is the computational power available. With enough power, the detector becomes perfectly reliable with no false alarms.
	
	In the future, with more computational power but also potentially more efficient solvers, attacks based on detecting incompatibilities can become more and more popular. In particular, it could be applied to other data formats such as H26x, HEIC, or any compression standard that uses a DCT method with quantization. Another approach would be to create a steganographic scheme in order to remain compatible with the DCT compression. This was already tried in~\cite{butora2022fighting} by reducing the variance of the spatial rounding error to fool the detector and we now know that the variance can be seen as a proxy to the compatibility property.
	
	\section*{Acknowledgments}
	\noindent The work presented in this paper received funding from the European Union’s Horizon 2020 research and innovation program under grant agreement No 101021687 (project “UNCOVER”). We also thank Tomas Pevny (Technical University of Prague) for pointing us to the use of solvers like Gurobi.
	
	\appendix
	\subsection{Discussion on heuristic metric}
	\label{sec:appendix_heuristic_metric}
	The main idea of the heuristic search is to rank neighbors depending on the metric~\eqref{eq:metric} and then explore the "best" candidate. This metric is completely fine for selecting good candidates, however, the algorithm will need more iteration to converge if most of the candidates have the same metric. Equality cases are the worst-case scenario when using a priority queue as we do. If the main metric $g_{\mathbf{c}}$ creates an equality, we use this second metric:
	\begin{equation}
	g'_{\mathbf{c}}(\mathbf{\tilde{x}}) = \left\Vert  \max\left\{ \left\vert \mathbf{c} - \frac{f_{\text{DCT}}(\mathbf{\tilde{x}})}{\mathbf{Q}}\right\vert - 0.49, 0\right\}\right\Vert_1
	\end{equation}
	Where the division and the max function are element-wise operations. We recall from equation (6) that if the distance $\left\vert \mathbf{c} - \frac{f_{\text{DCT}}(\mathbf{\tilde{x}})}{\mathbf{Q}}\right\vert$ is less than 0.5 for each component, then the pixel block $\mathbf{\tilde{x}}$ is an antecedent of the DCT block $\mathbf{c}$.

	This alternative metric is a mix between a norm infinity and a $\ell1$ norm in the sense that, we only use a subset of the components and sum their absolute values to calculate the distance. Every component already inside the block is ignored (the max function sets their value to 0), but we compute the $\ell1$ norm on the remaining values outside of the block.
	We used 0.49 instead of 0.5 because the borders are not always considered inside the block (it depends on the sign of $\mathbf{d}$ which is unknown).
	Finally, this metric is not indispensable. Even without it, the algorithm will work but could require more iteration to find an antecedent.
	
	This metric is particularly needed for \texttt{islow} and other integer DCT transforms. Indeed, candidates obtained with \texttt{islow} have quantized coordinate that makes them more likely to have the same metric. 
	
	This issue with quantized DCT transform explains the difference of probabilities between the naive DCT and the \texttt{islow} DCT visible in Fig.~\ref{fig:proba}. The true probabilities are probably very close to each other since both DCTs should yield almost the same result, but the heuristic algorithm is not as good when there are too many ties between different candidates and therefore is less likely to solve compatible blocks.
	
	\bibliographystyle{plain}
	\bibliography{incompatibility.bib}

\begin{thebibliography}{10}

\bibitem{agarwal2020photo}
Shruti Agarwal and Hany Farid.
\newblock Photo forensics from rounding artifacts.
\newblock In {\em Proceedings of the 2020 ACM Workshop on Information Hiding
  and Multimedia Security}, pages 103--114, 2020.

\bibitem{bas2011break}
Patrick Bas, Tom{\'a}{\v{s}} Filler, and Tom{\'a}{\v{s}} Pevn{\`y}.
\newblock ” break our steganographic system”: the ins and outs of
  organizing boss.
\newblock In {\em International workshop on information hiding}, pages 59--70.
  Springer, 2011.

\bibitem{benes2022knowyl}
Martin Benes, Nora Hofer, and Rainer B{\"o}hme.
\newblock Know your library: How the libjpeg version influences compression and
  decompression results.
\newblock {\em Proceedings of the 2022 ACM Workshop on Information Hiding and
  Multimedia Security}, 2022.

\bibitem{srnet}
Mehdi Boroumand, Mo~Chen, and Jessica Fridrich.
\newblock Deep residual network for steganalysis of digital images.
\newblock {\em IEEE Transactions on Information Forensics and Security},
  14(5):1181--1193, 2019.

\bibitem{butora2022fighting}
Jan Butora and Patrick Bas.
\newblock Fighting the reverse {JPEG} compatibility attack: Pick your side.

\bibitem{but23ih}
Jan Butora, Patrick Bas, and R\'{e}mi Cogranne.
\newblock Analysis and mitigation of the false alarms of the reverse jpeg
  compatibility attack.
\newblock In {\em Proceedings of the 2023 ACM Workshop on Information Hiding
  and Multimedia Security}, IH\&MMSec '23, page 59–66, New York, NY, USA,
  2023. Association for Computing Machinery.

\bibitem{butora2019reverse}
Jan Butora and Jessica Fridrich.
\newblock Reverse jpeg compatibility attack.
\newblock {\em IEEE Transactions on Information Forensics and Security},
  15:1444--1454, 2019.

\bibitem{cogranne2019alaska}
Rémi Cogranne, Quentin Giboulot, and Patrick Bas.
\newblock The {ALASKA} steganalysis challenge: A first step towards
  steganalysis.
\newblock In {\em Proceedings of the {ACM} Workshop on Information Hiding and
  Multimedia Security}, {IH}\&amp;{MMSec}'19, pages 125--137. Association for
  Computing Machinery.

\bibitem{denemark2014selection}
Tomáš Denemark, Vahid Sedighi, Vojtěch Holub, Rémi Cogranne, and Jessica
  Fridrich.
\newblock Selection-channel-aware rich model for steganalysis of digital
  images.

\bibitem{fridrich2001steganalysis}
Jessica Fridrich, Miroslav Goljan, and Rui Du.
\newblock Steganalysis based on jpeg compatibility.
\newblock In {\em Multimedia Systems and Applications IV}, volume 4518, pages
  275--280. SPIE, 2001.

\bibitem{guo2015using}
Linjie Guo, Jiangqun Ni, Wenkang Su, Chengpei Tang, and Yun-Qing Shi.
\newblock Using statistical image model for jpeg steganography: Uniform
  embedding revisited.
\newblock {\em IEEE Transactions on Information Forensics and Security},
  10(12):2669--2680, 2015.

\bibitem{hart_formal_1968}
Peter~E. Hart, Nils~J. Nilsson, and Bertram Raphael.
\newblock A formal basis for the heuristic determination of minimum cost paths.
\newblock 4(2):100--107.
\newblock Conference Name: {IEEE} Transactions on Systems Science and
  Cybernetics.

\bibitem{holub2014universal}
Vojt{\v{e}}ch Holub, Jessica Fridrich, and Tom{\'a}{\v{s}} Denemark.
\newblock Universal distortion function for steganography in an arbitrary
  domain.
\newblock {\em EURASIP Journal on Information Security}, 2014(1):1--13, 2014.

\bibitem{libjpeg}
IJG.
\newblock libjpeg.
\newblock \url{https://libjpeg.sourceforge.net/}, 1990.

\bibitem{kirchner2009synthesis}
Matthias Kirchner and Rainer B{\"o}hme.
\newblock Synthesis of color filter array pattern in digital images.
\newblock In {\em Media forensics and security}, volume 7254, pages 191--204.
  SPIE, 2009.

\bibitem{kodovsky2010quantitative}
Jan Kodovsk{\`y} and Jessica Fridrich.
\newblock Quantitative structural steganalysis of jsteg.
\newblock {\em IEEE Transactions on Information Forensics and Security},
  5(4):681--693, 2010.

\bibitem{levecque:hal-04098582}
Etienne Levecque, Patrick Bas, and Jan Butora.
\newblock {Compatibility and Timing Attacks for JPEG Steganalysis}.
\newblock In {\em {Workshop on Information Hiding and Multimedia Security}},
  Chicago, United States, June 2023. {ACM}.

\bibitem{pennebaker_jpeg_1992}
W.~B. Pennebaker and Joan~L. Mitchell.
\newblock {JPEG}: Still image data compression standard.

\bibitem{schlogl2023causes}
Alexander Schl{\"o}gl, Nora Hofer, and Rainer B{\"o}hme.
\newblock Causes and effects of unanticipated numerical deviations in neural
  network inference frameworks.
\newblock In {\em Thirty-seventh Conference on Neural Information Processing
  Systems}, 2023.

\bibitem{vazquez2013set}
David V{\'a}zquez-Pad{\'\i}n, Pedro Comesa{\~n}a, and Fernando
  P{\'e}rez-Gonz{\'a}lez.
\newblock Set-membership identification of resampled signals.
\newblock In {\em 2013 IEEE International Workshop on Information Forensics and
  Security (WIFS)}, pages 150--155. IEEE, 2013.

\bibitem{westfeld1999attacks}
Andreas Westfeld and Andreas Pfitzmann.
\newblock Attacks on steganographic systems: Breaking the steganographic
  utilities ezstego, jsteg, steganos, and s-tools-and some lessons learned.
\newblock In {\em International workshop on information hiding}, pages 61--76.
  Springer, 1999.

\end{thebibliography}
	
\end{document}